# Implementing Large-Scale Agile Frameworks: Challenges and Recommendations

Kieran Conboy & Noel Carroll

*Lero – the Irish Software Research Centre*
*National University of Ireland Galway, Galway, Ireland*
kieran.conboy@nuigalway.ie ; noel.carroll@nuigalway.ie

//Based on 13 agile transformation cases over 15 years, this article identifies nine challenges associated with implementing SAFe, Scrum-at-Scale, Spotify, LeSS, Nexus, and other mixed or customised large-scale agile frameworks. These challenges should be considered by organizations aspiring to pursue a large-scale agile strategy. This article also provides recommendations for practitioners and agile researchers.//

Large-scale agile development is increasingly prevalent in contemporary software development organisations. While there are many potential benefits, large-scale transformations are fraught with challenges such as communication issues, a lack of flexibility, and co-ordination challenges.

To address these issues, many have turned to large-scale agile development frameworks such as the Scaled Agile Framework (SAFe) [1], Large Scale Scrum (LeSS) [2,3], Spotify [4], Nexus [5], and Scrum at Scale [6]. Each incorporates predefined workflow patterns and routines, and is supported by an ever-increasing set of tools. However, empirical evidence regarding the adoption of such frameworks (Fig. 1), their use, effectiveness, and challenges is still very much in its infancy. This paper aims to address this.

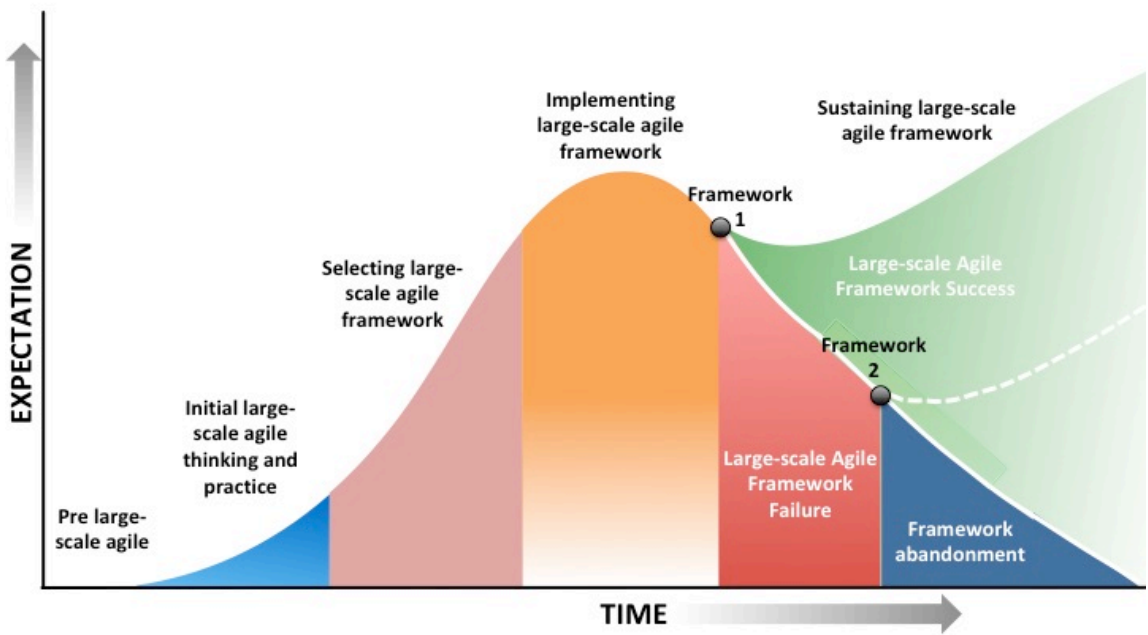

Figure 1: Adoption of Large-Scale Agile Frameworks

# A 15 Year Retrospective

We draw on 15 years of research collaboration experience with global companies, and in particular those listed below (Table 1). We have witnessed these organisations go through all parts of adopting frameworks (Fig.1) from the first tentative steps on their agile journey, through to the selecting, implementing and sustaining of a large-scale agile framework, combination of frameworks, or their own custom-built framework.

| *Organisation* | *Product/service* | *Years Studied* | *Scale (no. developers)* | *Framework(s) used* | *Team locations* |
|---|---|---|---|---|---|
| Accenture | Global management consulting, technology services, and outsourcing | 2001-2004 | 180 to 2,500 across >60 large-scale agile engagements | Mixed/custom frameworks | 42 sites- Europe, U.S., Asia, Australia |
| IrishBank* | Financial services | 2014-2018 | >150 | Mixed/custom frameworks | Ireland, UK |
| TechCo* | Products for software development and project management | 2013 | >340 | SAFe | Australia, Europe, San Francisco |
| Dell | Technology solutions services and support | 2014-2018 | >180 | Scrum at Scale | Ireland, U.S., India, China |
| FinanceCo* | Financial services | 2006-2018 | >1250 | Spotify, custom frameworks | Ireland, U.S., India, China |
| Information Mosaic | Global provider of post-trade securities and corporate actions processing solutions | 2010-2018 | >200 | Mixed/custom frameworks | Ireland |
| SemiCo* | Industrial engineering systems for power, and distribution | 2012-2018 | >1500 | LeSS | Ireland, U.S., China |
| Ericsson | Network and communication technology innovation and services | 2013-2015 | >200 | Scrum at Scale | Ireland, Sweden |
| Intel | Technology solutions services and support | 2005-2008 | >250 | Mixed/custom frameworks | Ireland, U.S. |
| ConsultingCo* | Management consulting, technology services | 2001-2018 | 60 to 1,100 across multiple large-scale agile consulting engagements | Mixed/custom frameworks | >50 sites- Europe, U.S., Asia, Australia, South America |
| RevenueCo* | Public sector - tax/customs authority for EU country | 2012-2018 | >200 | Spotify, SAFe | Anonymous |
| Rovsing | Software for space and satellite testing and simulation systems | 2007-2011 | 65 | Mixed/custom frameworks | Denmark |
| BankCo* | Financial services | 2010-2014 | >150 | LeSS | Australia |

**Table 1: Cases studied**

---

* Pseudonym to protect anonymity

In some cases the framework adoption has been a huge success (Framework 1 in Fig. 1), establishing some organisations as global thought and practice leaders in large-scale development. In others, the framework was not as successful and some organizations tried follow-up alternative frameworks ('Framework 2' in Fig. 1), or abandoned them completely. We have developed challenges and recommendations based on in-depth and long-term observation, interviews and on-going access to data, documentation and tools.

## Challenges of Large-Scale Agile Frameworks

Based on our industry research collaboration across 13 large-scale agile implementation projects, we identified nine of the most impactful challenges and a set of recommendations (Table 2) to mitigate each. Challenges and recommendations were only included where there was clear substantiated evidence to support their inclusion were identified in at least two case studies. However, to protect anonymity, the specific cases underpinning each challenge and recommendation are not listed.

### 1. Defining Large-Scale Agile Framework Concepts and Terms

Publications that launched frameworks such as SAFe and Spotify explain the basics very well, but once one applies them outside of their intended context of a specific framework, you quickly run out of guidance. Many developers talked about misunderstanding of framework concepts and routines, and large-scale inconsistencies in the way the framework was interpreted and applied in each case. Some showed that when 'abstract terminology is used', and there is a lack of thorough explanation, then subtle social and cultural nuances of agile get lost. This makes it extremely challenging for organizations examining key terminology in for example, LeSS or Scrum at Scale and their potential suitability for organisational-specific requirements. Some even question if certain frameworks such as Spotify and SAFe really have enough substance to be considered a framework or method. There were many cases where inconsistent meaning and interpretation were problematic. Team level inconsistencies can be ironed out quickly but differences across a large swathe of teams "get ingrained and the differences grow and fester." Also, as one transformation consultant suggested:

> "in the absence of something clear and definitive people just continue to do the same thing they always did."

All of our research participants agreed that there is a problem is defining large-scale agile. However, defining large-scale is of less a priority to some practitioners, as more emphasis in placed on the value that a framework can deliver. As one stated:

> "It doesn't matter what you call it – as long as the practices are adding value and reducing delivery times – it is good".

### 2. Comparing and Contrasting Large-Scale Agile Frameworks

Choosing between SAFe, LeSS, or Scrum at Scale was problematic for many organisations. Many noted the lack of any assessment model for conducting such a comparison to guide critical decisions on adopting specific large-scale agile frameworks. Some transformation leaders were required to justify their choice of frameworks, and so the absence of a comparison model was highly problematic, and in some cases stalled the agile transformation. As one agile champion noted:

> "I'm not sure management care whether we use LeSS, Scrum at Scale, or anything else, but they do need to know we reflected, evaluated options and justified our choice before they fund us."

Framework selection was often ad hoc where one or two read a book, attended a talk and the decision was made. Sometimes the decision cannot even be traced to source. One developer captured this sentiment:
> "One day we were doing Spotify…. but tight regulation and compliance mean we're not like them. We are not a music company. We should have thought about Large-Scale Scrum, SAFe, agile portfolio management or something else."

**3. Readiness and Appetite for Large-Scale Agile Frameworks**
Agile transformation requires staff and structures to be ready and willing to transform. However, staff can be ready for change software practices but not necessarily ready and willing to adopt a particular framework. For example, some organisations we have worked with have undergone multiple framework changes. One set of teams over a three year period went from being organised by 'service lines', to a set of Scrum teams in a Scrum of Scrums structure, to a 'full blown' SAFe implementation, and now to Spotify tribes and chapters. There was evidence that as frameworks are implemented, particularly multiple times, then either development teams get frustrated and may oppose framework adoption, or more commonly just take a passive approach and not do things any differently.

Also, explanations of large-scale frameworks such as SAFe and Scrum at Scale tend to explain their associated structures and processes, but provide little guidance on how organizations can assess their overall readiness or appetite to undertake a large-scale agile transformation process. Executives may sense a lack of readiness amongst some developers or groups, but don't have any mechanism to clearly identify these issues.

**4. Balancing Organisational Structure and Large-Scale Agile Frameworks**
Fitting a 'one size fits all' large-scale agile framework within an existing organisational structure is challenging for two reasons. First, large-scale frameworks come with pre-defined structures, routines and tools of their own, as opposed to a general custom approach. Second, these organisational structures are in constant flux in response to their external competitive or heavily regulated environments. Agile at a local level allows teams to fit into small, flexible, and dynamic teams and allows organizations to adopt to such structures in a flexible manner. Large-scale frameworks such as SAFe are much more dominant and small realignments can cause significant disruption across units of an organisation. Implementing and then maintaining a large-scale agile framework with an evolving organisational structure can be difficult when dealing with industry, e.g. organisation wide compliance processes or regulatory changes. In addition large-scale frameworks can often require changing organisational structures, which can be very challenging to do.

**5. Top-down versus Bottom-up Implementation of Large-Scale Agile Frameworks**
Many implementations were driven completely from either the bottom-up or from the top-down, rather than a mix. While bottom up implementation is well known to be most effective for 'small' Scrum implementation, it is not that clear in large-scale frameworks such as SAFe where senior management support and involvement is key to success. These were often absent in bottom-up implementations, and what resulted was often mass confusion as each 'tribe' or team drove different variants of the parent model, creating a fragmented mess of practices and expectations of other teams. Top-down implementations also had mixed success with many feeling this was yet another framework imposed by those who didn't understand the implications or problems to be solved, and didn't provide a clear path to implementation. As one stated:
> *"The very people imposing it still require the old reports and five-year plans that SAFe is supposed to eliminate".*

This is exacerbated by a lack of high quality training courses and coaching specific to large-scale agile frameworks, particularly at the executive and project portfolio level.
> *"There are Scrum coaches everywhere. But it is hard to find quality executive level SAFe coaching. Mostly, we get people that, while good, are venturing into world of SAFe for the first time."*

**6. Over-emphasis on 100% Framework Adherence over Value**
When a formal framework such as SAFe, Scrum at Scale or Spotify is used, there is a tendency to measure agile transformation by adherence to that framework, rather than the value it provides. For example, progress was often described in terms of the number of tribes established, or the number of teams participating in Scrum of Scrums. Ironically, quite a few noted that management place more emphasis on adherence to agile frameworks, rather than the impact on key performance metrics, primarily due to (i) the ability to immediately and tangibly measure the former, and (ii) difficulties in defining exactly how much a metric change was due to the framework versus other factors.

In almost all cases, the final 5-20% of development activity where the large-scale framework floundered or caused significant problems, and very often 80% of the stress and effort expended went into achieving the last 5%. It was clear that such endless pursuit towards a 100% transformation just didn't make sense. For example, one particular company required FDA compliance certificates that took two to three years to approve. However, managers have little guidance on how to find the optimal degree of transformation.

**7. Lack of Evidence-Based Use of Large-Scale Agile Frameworks**
While foundation papers to SAFe, Scrum, LeSS and Scrum at Scale exist, there is a lack of empirical case studies which subsequently apply those frameworks 'in the wild'. There were many examples of 'brick walls' across organizations – instances where staff claimed there was a particularly difficult problem or contextual issue for which the original framework papers had no guidance. These included heavily regulated industries or products, software that required more research-intensive work than development, and one case involving a small number of very niche, specialised developers who had to split their time across 450 projects. Participants were particularly frustrated at a lack of cumulative tradition empirical studies which do not build on the original framework papers, but rather focus on next generation application of large-scale agile developments and revised ways of 'doing agile'.

**8. Maintaining Developer Autonomy in Large-Scale Agile Frameworks**
Today's developers expect and often demand autonomy in how they work. Remote working, flexible hours, 'bring your own devices', devolving work to crowd platforms, various instant messaging and media platforms, and a plethora of niche tools and apps were common across most of the cases studied. Also, autonomy to tailor and improvise how they work was always facilitated by traditional agile methods.

It is already known that autonomy becomes increasingly difficult at scale. However, large-scale agile frameworks exacerbate this problem, and impose even more restrictions and rigidity. There were many instances where developers requests to implement processes and tools were no longer accepted as they were not viewed as compliant with the new SAFe implementation. Some referred to Spotify's tribes structure "even dictating where we are allowed to sit". As one participant stated:

*"Autonomy and flexibility is what agile is all about. But when employees want so many different things, and we allow them all, then the SAFe carcass gets slivered away to such a degree there is nothing recognisable left."*

**9. Misalignment between Customer Processes and Large-Scale Agile Frameworks**
Organisations are now expected to include customers into their process design and in many cases are encouraged to completely align with customer processes in a seamless manner. However large-scale agile frameworks were more challenging as their pre-defined practices and structures are harder to hide and subtly relabel. Customers struggle with some terminology. In fact, to some that are not familiar, the term 'SAFe' invokes connotations of heavyweight regulatory and compliance processes, rather than anything light and nimble. This meant that most organisations had to drop the framework for that piece of work, or else blend them in some way. This becomes incredibly challenging as some organisations studied, for example, with hundreds of clients ranging from small to big enterprises, small to big revenue, and massive diversity in terms of development and reporting frameworks. In some cases, organisations needed to form agreements on how to work around a specific large-scale agile framework to collaborate on and deliver software products at certain points through increased customer involvement.

| Challenge | Recommendation |
|---|---|
| 1. Defining Large-Scale Agile Framework Concepts and Terms | • Spend time reflecting and defining what is meant mean by 'agile' and 'scale' in your organisational context before adopting a large-scale agile framework.<br>• Establish clear motivation to scale agile development to meet business needs.<br>• Develop a common vocabulary to capture vision and value of large-scale agile transformation.<br>• Ensure the common vocabulary is accessible, coherent, and promoted across all stakeholders in the early stages of adopting a large-scale agile framework. |
| 2. Comparing and Contrasting Large-Scale Agile Frameworks | • Avoid comparison against methods (i) out of context or (ii) without framework-independent criteria to meet organisational-specific requirements.<br>• Use metrics that are core to an organisation's value portfolio to evaluate how each framework contributes to organisational productivity and performance for example, employee engagement, customers' satisfaction, productivity, agility, time-to-market, or quality.<br>• Use a small number of metrics (one to four), aggregated from other metrics if necessary to compare and contrast large-scale agile frameworks.<br>• Ensure comparison and justification of framework selection is clear to all key stakeholders in the large-scale agile transformation process. |
| 3. Readiness and Appetite for Large-Scale Agile Frameworks | • Conduct an organisational readiness assessment to examine potential barriers of adopting specific large-scale agile transformation frameworks.<br>• Identify gaps/issues and associated steps to resolve issues e.g. increased training, organisational structural changes, or new management styles or strategies.<br>• Use an incremental adoption of a large-scale agile framework in areas of weakness to ensure a smooth transformation process and demonstrate 'small wins'. |
| 4. Balancing Organisational Structure and Large-Scale Agile Frameworks | • Identify what new structural requirements a specific large-scale agile framework imposes on an organisation.<br>• Evaluate how new agile framework structures will positively or negatively impact on performance, standards compliance, and flexibility across an organisation. |
| | • Weigh-up the benefits and drawbacks of the large-scale agile framework and how they may alter business operations. |
| 5. 'Top-Down' versus 'Bottom-Up' Implementation of Large-Scale Agile Frameworks | • Determine whether a large-scale agile framework promotes a top-down or bottom-up implementation approach.<br>• Strike a clear balance between enabling top-down and bottom-up transformation.<br>• Provide continuous education or training opportunities at all staff levels including executives, project leaders, and software developers.<br>• Continually support and reflect on implementation activities from top and bottom. |
| 6. Over-Emphasis on 100% Framework Adherence over Value | • Determine whether the organisations agile transformation prioritises adherence to specific agile frameworks or whether the overall success of the method is better for business.<br>• Identify which transformational factors will influence adherence over value, such as standards compliance, speed, cost, technology, or customer requirements.<br>• Plan for the optimal degree of transformation with the large-scale agile framework as per your organisational goals and objectives. |
| 7. Lack of Evidence-Based Use of Large-Scale Agile Frameworks | • Build evidence (e.g. metrics) to support the use of a particular large-scale agile framework to transform your organisation.<br>• Regularly test scalability at a more sustainable pace to learn your way through the transformation process, e.g. through transformational "small wins".<br>• Identify and contextualise issue to offer guidance on agile large-scale transformation and establish best practice. |
| 8. Maintaining Developer Autonomy in Large-Scale Agile Frameworks | • Engage with the people to assess their overall satisfaction in relation to autonomy in the workplace provided by the large-scale agile framework.<br>• Carry out regular audits to ensure awareness and adaptation of a large-scale agile framework remains transparent and relevant within and across projects and teams.<br>• Explore whether new policies, such as 'bring your own device' would improve autonomy and facilitate a smooth large-scale agile transformation. |
| 9. Misalignment between Customer Processes and Large-Scale Agile Frameworks | • Consider involving customer stakeholders during the selection of a large-scale agile framework to increase transparency, cooperation, and alignment.<br>• Examine how the choice of a large-scale agile framework will provide the organisations with some flexibility to cater for growing dynamic customer needs. |

**Table 2: Summary of challenges and recommendations**

> # RELATED WORK ON LARGE-SCALE AGILE FRAMEWORKS
>
> There are various publications describing available frameworks e.g. SAFe, LeSS, Spotify, Nexus, and Scrum-at-Scale. However, there is very little empirical research examining the common challenges associated across the range of large-scale agile frameworks.
>
> Dikert and colleagues [1] present a systematic literature review on large-scale agile transformations outlining challenges and success factors. Their results also describe various agile frameworks but demonstrate the lack of scientific studies that focus directly on the transformation process.
>
> Studies such as Power [2] and Rolland et al. [3] show the lack of consensus and evidence underpinning any particular agile framework, and that the entire organisation does not need to become agile.
>
> Within the literature, there are many discrepancies in defining both 'agile' and 'large-scale transformations' [4] and distinctions between 'agile approaches' and 'organisational agility' [2].
>
> Previous research has explored specific challenges of large-scale agile development such as team co-ordination [5], and the threat to self-organisation as agile is scaled [6]. In addition, researchers have presented some success factors and recommendations for large-scale agile development. However, all call for more empirical evidence on agile within large-scale settings.
>
> **References**
> 1. K. Dikert, M. Paasivaara, and C. Lassenius. "Challenges and success factors for large-scale agile transformations: A systematic literature review." Journal of Systems and Software, Vol. 119, pp. 87-108, Sept. 2016
> 2. K. Power. "A model for understanding when scaling agile is appropriate in large organizations." In International Conference on Agile Software Development, 2014. pp. 83-92.
> 3. K. H. Rolland, B. Fitzgerald, T. Dingsoyr, and K-J Stol. "Problematizing agile in the large: alternative assumptions for large-scale agile development." In Proc. International Conference on Information Systems, Dublin, pp. 1-21.
> 4. M. Kalenda, P. Hyna, and B. Rossi. "Scaling agile in large organizations: Practices, challenges, and success factors." Journal of Software: Evolution and Process, vol. 1954, 2018. Volume 30, Issue 10, pp. 1-25
> 5. M. Paasivaara, C. Lassenius, and V. T. Heikkilä. "Inter-team coordination in large-scale globally distributed scrum: Do scrum-of-scrums really work?." In Proceedings of the ACM-IEEE international symposium on Empirical software engineering and measurement, 2012, pp. 235-238
> 6. N.B. Moe, D. Šmite, A. Šāblis, A-L. Börjesson, and P. Andréasson. "Networking in a large-scale distributed agile project." In Proceedings of the 8th ACM/IEEE International Symposium on Empirical Software Engineering and Measurement, 2014, p. 12

## Conclusions

Drawing on 15 years experience across 13 cases, this paper identifies nine challenges associated with the implementation of large-scale agile frameworks. Organisations considering, planning or in the midst of agile transformation can use our study to identify and pre-empt challenges they may be particularly susceptible to. Such an exercise can be insightful, given that many problems are subtle and can exist 'under the radar'. We would particularly encourage organisations to take a multi-layered approach across different employee and stakeholder groups. In terms of limitations, while reflection can expose challenges, their complete removal may be difficult. Also, it may not be feasible to implement all recommendations due to restrictions such as cost, culture, structure and span of control.

## Acknowledgement

This work was supported with the financial support of the Science Foundation Ireland grant 13/RC/2094.

## References

1. D. Leffingwell, 2015. Scaled agile framework. Available online (20/07/18): http://scaledagileframework.com/

## Author Bios

**Kieran Conboy** is a Professor at NUI Galway, and previously worked for Accenture Consulting. Kieran has published over 200 articles in leading journals including ISR, EJIS, JAIS and ISJ. His research examines contemporary technology management and design including concepts such as temporality, flow, open innovation and agility. He is editor of the European Journal of Information Systems and has chaired many international conferences in his field.
*Contact him at: kieran.conboy@nuigalway.ie*

**Noel Carroll** is a lecturer of information systems at NUI Galway, and works with Lero – The Irish Software Research Centre. His research interests include seeking ways to support organizations in developing large-scale transformation strategies. He has edited special issues, published, chaired, and reviewed for leading international journals and conferences in his field.
*Contact him at noel.carroll@nuigalway.ie*